# Teaching Model-based Requirements Engineering to Industry Professionals: An Experience Report

Marian Daun*, Jennifer Brings*, Marcel Goger†, Walter Koch† and Thorsten Weyer*

*paluno - The Ruhr Institute for Software Technology
University of Duisburg-Essen, Essen, Germany
{marian.daun, jennifer.brings, thorsten.weyer}@paluno.uni-due.de

†Schaeffler AG, Herzogenaurach, Germany
{marcel.goger, walter.koch}@schaeffler.com

*Abstract*—The use of conceptual models to foster requirements engineering has been proposed and evaluated as beneficial for several decades. For instance, goal-oriented requirements engineering or the specification of scenarios are commonly done using conceptual models. Bringing such model-based requirements engineering approaches into industrial practice typically requires industrial training. In this paper, we report lessons learned from a training program for teaching industry professionals model-based requirements engineering. Particularly, we as educators and learners report experiences from designing the training program, conducting the actual training, and applying the instructed material in our day-to-day work. From these findings we provide guidelines for educators designing requirements engineering courses for industry professionals.

## I. INTRODUCTION

In requirements engineering research, graphical models are commonly used. Not only are model-based approaches to support requirements engineering regularly proposed, but, furthermore, approaches commonly use models for documentation and analysis purposes. Model-based approaches deal, for instance, with context analysis (e.g., [1]), goal-orientation (e.g., [2]), scenarios (e.g., [3]), or a more detailed analysis of potential solutions (e.g., [4]). Furthermore, model use is regularly proposed to foster elicitation (e.g., [5]), analysis (e.g., [6]), and negotiation (e.g., [7]) of requirements.

Among the benefits of model orientation, the support of traceability to later artifacts is often valued (e.g., [8]). This allows, for instance, to directly link architecture decisions to design rationales [9]. Other benefits are seen in the support for test case generation [10] or in the support for safety analyses [11]. Beside these requirements engineering specific benefits, general benefits also exist, like coping with the growing complexity of systems, the growing demand for safely functioning software, or the potential for re-use [12].

While the benefits of model-orientation are well-known and approaches are proposed on a regular basis in requirements engineering research, there is a gap when it comes to the introduction of these techniques into industrial practice [13]. To foster the introduction of these requirements engineering practices into industrial practice, different kinds of approaches for technology transfer exist (e.g., [14]). Commonly industrial training is seen as either a precondition for industry application (e.g., [15]) or proposed as an approach on its own (e.g., [16]).

In this paper, we report on our (i.e., educators and teachers) insights gained from designing and applying a training program to bring model-based engineering into practice. The development of a training program is part of a larger endeavor to foster seamless model-based engineering where multiple industry partners and academia are involved.[1] In this paper, we place emphasis on a course for model-based requirements engineering and its application at the Schaeffler AG. We report findings from the perspective of educators designing the training program, from the perspective of educators and industry professionals conducting the model-based requirements engineering course together, and from the perspective of industry professionals applying the instructed material in our day-to-day work. From our findings and under consideration of findings from other researchers, we synthesize guidelines for educators. This includes the identification of pitfalls as well as hints for designing a course setup that meets industry needs.

The paper is outlined as follows: Section II gives background information on the course design, the participating learners, and the application of the instructed material in learners' day-to-day work. Section III contributes the major findings and experiences gained. Section IV gives an overview over the related work. Section V discusses the findings and synthesizes guidelines. Finally, Section VI concludes the paper.

## II. COURSE DESIGN

In the following subsections, we will briefly introduce the course. In particular, we will discuss our goals in teaching model-based requirements engineering and the instructed material. For a detailed discussion about the course setup and the teaching approach used please refer to [17], where we present the course design and provide initial results from the application in a university setting.

---
[1]This research is part of the SPEDiT project https://spedit.informatik.tu-muenchen.de. Which aims at transferring research results from its predecessor projects SPES and SPES XT http://spes2020.informatik.tu-muenchen.de into industrial practice.

## A. Teaching Goals

The course aims at teaching model-based requirements engineering. The main goals of the course are model creation, model understanding, model evolution, and model verification:

*Model creation.* After taking the course learners shall be able to create requirements models from scratch. Therefore, learners need to gain a thorough understanding of the modeling languages used, including their syntax and semantics, as learners shall be enabled to create requirements specifications for real systems at an advanced level of complexity.

*Model understanding.* Learners shall be able to interpret requirements models. This in particular means that the learners must be able to read, understand, and correctly interpret requirements models, which have been created by others. Furthermore, learners shall not only be able to understand requirements models on their own, but also the relations between different requirements models.

*Model evolution.* Learners must understand that requirements change during development projects and that it is necessary (or at least advisable) to update the requirements specifications accordingly. Hence, learners shall be able to revise existing requirements models. Learners shall also understand and be able to make use of the complementary nature of different models in requirements engineering, which leads to iteratively updating models.

*Model verification.* Learners shall understand the need for regular model inspections and quality assurance of requirements artifacts. Learners shall be able to check whether a certain model correctly represents a certain set of stakeholder intentions (cf. [18]). Furthermore, learners shall be able to check for consistency within requirements models and between requirements models and other engineering artifacts from later stages. When it comes to automated verification, learners shall understand the potential application and benefits as well as its shortcomings compared to manual model validation.

In addition, the course also places emphasis on general aspects of requirements engineering such as elicitation, negotiation, and traceability of requirements.

## B. Syllabus

To achieve the teaching goals four teaching blocks have been defined.

*Context.* The context of a system comprises phenomena, which are potential sources for requirements and therefore potentially have an impact on the requirements and on the system itself. Beside the properties of the operational environment (e.g. actors, external systems, and their relevant properties), the context also contains, for instance, standards and legal provisions which need to be considered [19]. In this teaching block, we differentiate between models of the operational environment and models of the context of knowledge [20]. When analyzing and modeling the operational environment, we apply three modeling perspectives (information structure, functional, and behavioral) [21]. Models of specific perspectives on the operational environment serve as a basis for (automated) validation and verification activities (e.g. behavioral models in simulation-based quality assurance). Models of the knowledge context are used to document the requirements sources, their relevant relationships among each other, and their impact on the requirements. In our training program, these context of knowledge models are important means for structuring the elicitation, refinement and validation process for requirements.

*Goal- oriented requirements engineering.* In goal-oriented requirements engineering [22] goals (i.e., objectives the system under development shall achieve) are defined on different levels of abstraction. This allows for starting with quite informal requirements that are very close to the originating stakeholders' ideas and refine them into more fine-grained detailed requirements. In the training program, we use the ITU Goal oriented Requirements Language (GRL) [23] to document goals as it has already proven useful in industrial settings [24]. GRL is a language to hierarchically structure goals and to define dependencies between them. GRL differentiates between different kinds of goals. In addition to hard and soft goals as used by most languages for goal modeling, tasks and resources are defined. Additionally, GRL allows differentiating between goals of different actors and analyzing their relations. The teaching block for goal modeling shall enable learners to create and evolve GRL goal models on their own.

Figure 1 gives an impression of the learning material on goal modeling. In a lecture-style introduction video, the modeling constructs of the GRL are introduced and a simple goal graph is shown. Emphasis is given to the structuring of the goals, the use of different kinds of decomposition links as well as contribution links. Also the principle of goal fulfillment (i.e. all sub goals must be fulfilled to ensure a decomposed goal is fulfilled, too) is introduced. The learning material places particular emphasis on discussing different alternatives for goal structuring depending on the specific purpose.

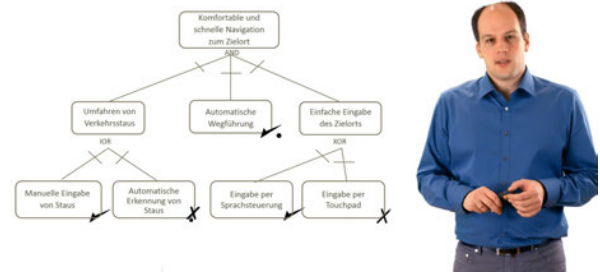

Fig. 1. Lecture-style introduction to ITU GRL goal models

*Scenarios.* Scenarios show exemplary executions of the system. For instance, scenarios can be used to capture how a stakeholder believes the system should work. In the engineering of embedded systems, scenarios often give particular em-phasis to the definition of the interactions exchanged between the system and its context or between pre-defined components of the system under development. ITU Message Sequence Charts (MSC) [25] is an interaction-based language to define the intended system behavior in terms of interaction sequences. MSC and comparable languages such as sequence diagrams or live sequence charts are commonly used in scenario-based requirements engineering approaches (e.g., [26]). The ITU recommendation defines two types of diagrams. Basic MSCs (bMSC) define the interaction-based behavior as intended by one scenario. Additionally, high-level MSCs (hMSC) are used to structure the scenarios in terms of their execution order. The teaching block for scenario modeling shall enable learners to create and evolve MSC scenario models on their own.

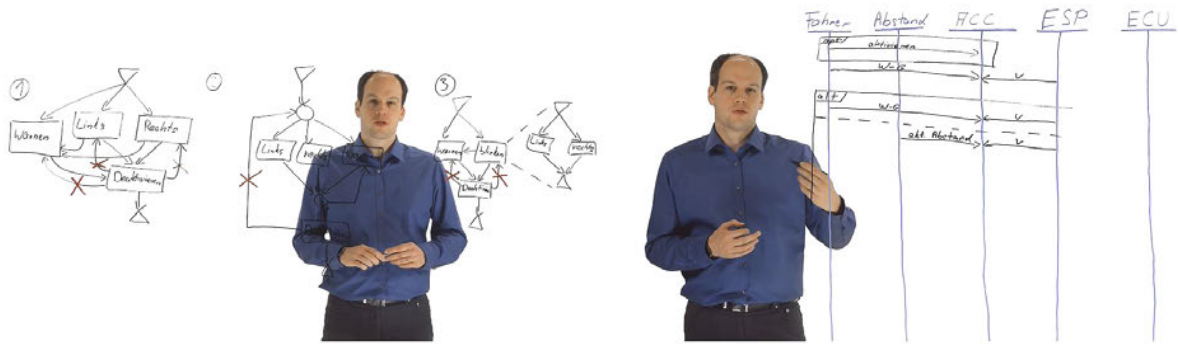

Fig. 2. Whiteboard-style solution discussion for ITU MSC scenario models (left: hMSC, right: bMSC)

Figure 2 gives two example screenshots from the learning material for scenario modeling with ITU MSC. As in the case of goal models, emphasis is given to discussing advantages and disadvantages of the different possibilities and allowing learners to gain an understanding on when best to use which structuring alternative. For bMSC different alternatives are also discussed, particularly under the aspect of hMSC structuring. It is shown that larger bMSC are better suited for discussing entire scenarios in a self-contained fashion, while at the same time the larger nature of these bMSC diagrams makes them harder to understand and decreases the scenario's usefulness in stakeholder discussions. On the other hand, keeping scenarios too brief will impede comprehensibility as important information is missing.

*Detailed System Requirements.* These are requirements in the sense of ISO/IEC/IEEE Std. 24765:2017 Definition 3.3431-2 [27]. System requirements always rely on a certain solution concept for the system to be built and are defined by means of context-controlled phenomena visible to the system and system-controlled phenomena visible to the environment [28]. Similar to the structured modeling of the operational environment, we teach three distinct specification perspectives for system requirements, which are closely integrated by mapping relationships to obtain a coherent specification model of the system requirements. Along with quality goals, system requirements are the major source of information for the architectural design and implementation of the system. As has been established (e.g., [21]), we use data models, function models, and behavior models to capture detailed system requirements. Figure 3 presents an excerpt from the learning material showing a behavior requirements model for an electronic stability control. In the learning material, we use UML state machine diagrams to capture the important global states of the system, the transitions between them, and the essential behavior the system shall exhibit in the individual states.

The modeling languages taught were chosen in coordination between educators, learners, and the learners' company to ensure appropriateness. One reason for choosing them was that they had already been used and adapted in prior projects and have established themselves as useful in other companies in the embedded domain.

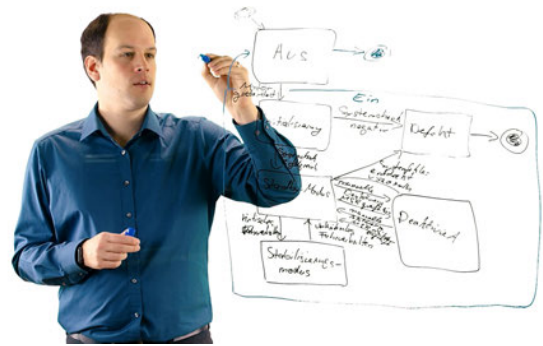

Fig. 3. Whiteboard-style introduction to the use of UML state machine diagrams for specifying detailed system requirements

### C. Participants

As a basis for the course setup and training concept, personas have been defined by the industry partners. The industry partners were asked to define typical characters with detailed backgrounds that should be considered as fictitious, potential participants of the trainings. At this time it was not yet defined in what project or business unit the pilot project would take place. In total five personas have been defined, each character taking a different role (i.e., system architect, requirements engineer, test engineer, software architect, and software developer) in the software development process.

TABLE I
BACKGROUND AND JOB DESCRIPTIONS OF ACTUAL PARTICIPANTS

|  | Participant 1 | Participant 2 | Participant 3 | Participant 4 | Participant 5 |
| --- | --- | --- | --- | --- | --- |
| Educational Background | Mechanical Engineering | Mechatronic Engineering | Electrical Energy & Automation Technology | Electrical engineering | Electrical engineering |
| Process role/ Job title | System Architect & Requirements Engineer | System Architect & System Designer | Software Requirements Engineer | Software Architect | Software Developer |

After one year of preparation time the pilot project was selected and started. The choice fell on a development project in the automotive e-mobility domain and included two departments – software and (mechatronic) system development. The real evaluation group, at the end consisted of approximately 20 people attending the course to get used to the basic principles of model-based engineering. Five people out of that group accomplished the trainings combined with 2-weekly coaching sessions over a period of 1.5 years. Table I shows the actual backgrounds of the actual participants in the pilot training program.

*D. Procedure*

The application of the training program consisted of two steps. First, industry professionals were taught model-based requirements engineering using the created learning material. Second, the instructed material were applied by the participants in industrial practice under supervision of the instructor.

*1) Teaching model-based requirements engineering in industry:* First, industry professionals are instructed with the material using the online course, so they can study at their own pace, dependent on the current demands of the development project. Then industry professionals build simple examples, model academic cases to get an understanding of the instructed material, and ask questions about them in the webinars.

*2) Application of the instructed material in industrial practice:* To apply their newly learned skills, our participants are developing a highly integrated parking lock actuator for vehicles with an electric axle drive. Even though the company has already developed several predecessor products, the development of this product was started from scratch to allow participants to experience the development of an entire product in model-based fashion and, hence, allowing the company to assess the value of model-based engineering. Participants prepare the required models and questions for workshop sessions where they receive feedback on their models and answers to their questions from the instructors. Furthermore, the models are improved by the participants and instructors together.

## III. RESULTS

This section discusses our findings from conducting the course and applying the instructed material in industrial practice. Table II reports each major finding in a structured manner by separating between symptom, cause, action taken and result. The symptoms detail our observations of aspects that needed improvement. Once identified, we discussed the matter among educators as well as with industry representatives and the participants to gain an understanding of the respective cause. After we identified the cause we designed a response and took action to improve the situation. Finally, we report the results we observed after the action taken. Based on these symptoms and causes identified and the results of the mitigating actions taken, we will - under consideration of findings from other researchers reported in the literature - define guidelines in Section V.

## IV. RELATED WORK

As findings form Section III rely on observations from one course and are therefore subject to threats to validity, this section introduces findings from other researchers and relates them to our original findings. This allows gaining a more objective and generalizable view on the introduced findings and allows drawing conclusions and formulating inferences within the next section.

*A. Educational Approaches for Technology Transfer*

Regarding the technology transfer from academia into industrial practice, several approaches have been proposed. Most of them have in common the idea that bringing technology into practice requires teaching and training of the new technology. Approaches either focus on the long-term transfer of knowledge or on the short-term transfer. Long-term means teaching students the new techniques. When students graduate and find their way into practice, more and more taught material will be available for industry. In this setting, students typically serve as change agents (e.g., [29], [30]). Often it is suggested that the teaching setup should be aligned between university and industry to ensure relevancy (e.g., [30]). Note that such approaches are seen as controversially by some researchers. For instance, Hallinan and Paul [31] have concluded that the impact individual student can have is negligible. For short-term technology transfer, industrial training is often proposed, i.e., education of professionals (e.g., [16]). These education programs not only provide industry with better skilled staff but also provide researcher insight into industrial problems, as we can also confirm from developing and implementing our training program in industrial practice.

*B. Online Education*

In this section, we place emphasis on related work dealing with the use of online courses as this was a prerequisite from

TABLE II
FINDINGS

| ID | | Description |
|---|---|---|
| 1 | **Symptom** | Learners struggled with requirements engineering practices, resulting in imprecise models. Particularly, the naming of elements was sometimes somewhat ambiguous or misleading. |
| | **Cause** | When designing the training program, it was decided to only teach the modeling aspects of model-based requirements engineering to ensure a reasonable duration for taking the training program. It was assumed that all participants were experts in the field of requirements engineering. As we observed that on the one hand participants sometimes struggled with certain aspects of requirements engineering practices, but on the other hand learners were obviously experts in the field, we investigated this matter in more depth. Our understanding is, that participants typically know what they are doing, how to document requirements as it was common in their company and that this worked pretty well. |
| | **Action taken** | Due to the habits the participants had formed, we decided to also teach good requirements engineering practices and general requirements engineering guidelines. |
| | **Result** | This more general approach to training model-based requirements engineering led to learners placing more emphasis on good requirements engineering practices. It supported learners in remembering why they had been doing the things they did in the first place and helped them transfer their knowledge to the new situation of model-based requirements engineering. |
| 2 | **Symptom** | Learners often inquired why things should be done a certain way (e.g., why a model should contain different diagrammatic representations instead of one complex one). This was also reflected in the initial models created, where such good practices were often neglected. |
| | **Cause** | The initial idea of the training program was to minimize the time needed by the learners and to focus on industrial training and not so much on lecture-like transfer of knowledge. |
| | **Action taken** | We placed considerably more emphasis on teaching the benefits of a certain approach or a best practice and on why learners should follow these guidelines, as had been requested by the learners. |
| | **Result** | Learners were more satisfied with the teaching materials and were more eager to try things the way they had been instructed to do. |
| 3 | **Symptom** | First conducted exercises and initial applications of the instructed materials to the case studies were error prone and lacked sufficient detail for documenting core requirements of the project. |
| | **Cause** | We observed that this was particularly the case when participants applied modeling languages they were already familiar with (e.g., message sequence charts for scenario modeling). It was revealed that this stemmed from the fact that learners used these languages to create models the way they were used to, neglecting the fact that they were now creating requirements models. |
| | **Action taken** | We revised the training materials and placed therein and in the webinars and workshop meetings emphasis on the ontological foundations of requirements models. Thereafter, we introduced the modeling languages and showed how they were used for requirements modeling. Particularly, training materials now explicitly stress the difference between using a certain modeling language in requirements engineering and in design. |
| | **Result** | Learners placed more emphasis on modeling the requirements aspects. This also means that sometimes models created by the learners were initially a bit incorrect regarding the concrete modeling language used, but the intention of the created models (i.e. documenting requirements of the system) improved considerably. |
| 4 | **Symptom** | The application of the instructed material to the learners' project work resulted in models that were either too detailed or provided too little information. |
| | **Cause** | Learners tended to document the solution they had in their mind right away, leading to very detailed models, but at the same time not documenting important constraints. |
| | **Action taken** | We integrated more materials on good general requirements engineering practices and placed emphasis on teaching the benefits of good requirements engineering. |
| | **Result** | Participants placed more emphasis on documenting all relevant information. |
| 5 | **Symptom** | Learners often inquired how to bring their "classic world" of natural language requirements together with the newly instructed material on model-based requirements engineering. |
| | **Cause** | In the design of the training program it was decided to strictly focus on model-based aspects as it was assumed that participants are aware of how to document natural language requirements and should therefore not waste time being taught what they already know. |
| | **Action taken** | We revised the material to also teach good practices for requirements engineering in general. Furthermore, we discussed the proper integration of natural language and models. It showed, that the use of goal models was perfect to trace one natural language requirement to one goal and vice versa. |
| | **Result** | Learners rethought their already existing natural language requirements and improved them where necessary. The link between goals and natural language requirements allowed learners to create models for their project work in a more structured way. |
| 6 | **Symptom** | Learners struggled with the relationship between the different models, which led to inconsistencies between different models (e.g., between the specified context and the context instances used for scenario modeling). It was often assumed that the different models were independent of each other or learners saw that there was a relation but could not grasp them exactly. |
| | **Cause** | In the learning materials, we incorporated material that discussed the relationship between different models (e.g., between scenarios and detailed requirements) on an abstract level and showed examples. We did not place emphasis on concrete rules and approaches for linking together different models. In designing the course this was agreed upon to avoid immediately burdening participants with too much in-depth material. |
| | **Action taken** | We incorporated training sessions for traceability management to make learners aware of this concept and we provided basic rules that made it obvious how elements of one model relate to elements of another model. |
| | **Result** | Teaching traceability concepts and exemplifying them for the different types of models (i.e. context, goals, scenarios, and detailed system requirements) aided the learners in understanding that there are relations and how these relations manifest themselves. This led to a higher consistency between different models created by the learners. |
| 7 | **Symptom** | The models created by the learners were often lacking an appropriate level of detail for requirements. |
| | **Cause** | The training program was designed to not only train the techniques but also the application of the techniques with the company's standard modeling tool. This tool, was not adequate for requirements engineering purposes as it often placed emphasis on modeling very fine-grained details. Additionally, as learners acted in their well-known environment, they did what they regularly did, forgetting they were supposed to be documenting requirements. |
| | **Action taken** | The participants actually recognized this problem themselves and countered it by using simple tools such as a simple drawing program, to be able to place more emphasis on the instructed material than on the tooling. Subsequently, learners transferred this when using the original tool in a second step. In the meantime, we provided a simpler modeling tool, which supports learners by providing shapes for the modeling languages used. |
| | **Result** | Placing emphasis on the instructed material in the first place, and on the tool in the second place (i.e. after the instructed material has been learned), resulted in better models and a faster progress in learning. |
| 8 | **Symptom** | The instructed material is not sufficiently internalized by the learners. |
| | **Cause** | Learners focusing on the application to their work environment, tend to overlook important aspects of the instructed material. Learners only focusing on the instructed material, not on the impact for their day-to-day work, tend to study too superficial. |
| | **Action taken** | To avoid this, the training program contains exercises emphasizing difficult and important aspects of the instructed material and, in addition, the instructed material is applied in a real development problem. |
| | **Result** | This allowed the learners to gain a deeper understanding of the instructed material and the ability to apply the instructed material right away. Additionally, it allowed the instructors to get a deeper understanding of industry needs. |

our industry partners. In recent years, the use of online courses and material has been widely suggested to improve university education (e.g., [32]) as well as industrial training (e.g., [33]). In particular, blended learning [34] and flipped classroom courses [35], [36] gained much interest. Online courses are often proposed for teaching programming and basic computer science education. The use of online courses is less common for teaching conceptual modeling. For one example, Sedrakyan et al. [37] propose using feedback-enabled simulation to increase the students' learning abilities in conceptual modeling of business requirements.

Experiences with online courses for teaching software engineering have been commonly reported. For instance, Weston and Quinn [38] report on a survey conducted among instructors who use online material from a digital collection for teaching basic computer science classes. They report that instructors experienced multiple benefits from the use of the material, e.g., more efficient learning and more student engagement. In addition to such general surveys, experience reports exist, reporting that the use of online courses or online material in classes aids the students' learning experience as well as their motivation (e.g., [39], [40]). Even more importantly, it is often shown that students' performance has improved (e.g., [41], [42]). Further advantages deal with the individualizability of online courses, students can follow the course in their own pace and decide which elements are supportive for them [32].

### C. Teaching Requirements Engineering

Research on requirements engineering education commonly deals with the use of certain techniques to foster students' understanding of the matter. Particularly, in requirements engineering education research, emphasis is given on how to ensure an almost realistic experience for university students. Therefore, for instance, the idea of bringing in real stakeholders has been proposed (e.g., [43]), or the idea to simulate a realistic requirements engineering setup for globally operating companies (e.g., [44]). While most requirements engineering education research focuses on teaching requirements engineering in a university setting, approaches for requirements engineering training in industry have also been proposed (e.g., [45]). For an in-depth overview over the field of requirements engineering education, please refer to a recently published systematic mapping study by Ouhbi et al. [46].

### D. Comparison of Findings with Related Work

The need to focus on the core aspects of requirements engineering is long known, e.g., [47]. However, this finding is not specific to model-based requirements engineering but is emphasized by requirements engineering literature in general. The concept of emphasizing the why is not only limited to teaching requirements engineering but also to requirements engineering itself, see e.g., [48]. The same holds for the care that must be taken to avoid over and under specifications in requirements engineering, e.g., [49]. The need for tool support in teaching setups has often been valued, including for online courses. For instance, Burgueño, Antonio & Gogolla [50] report on the benefits of tool support for teaching UML and OCL models. Particularly, they also highlight the need to consider relationships (i.e. traceability) between different views using different models. They conclude that for this task the availability of an adequate tool also benefits the teaching. Furthermore, going into the direction of our finding that less expressive tools can also be supportive in learning environments, the authors found that confronting students with multiple tools, students were able to easily adapt from one tool to another, even if the tools differed in expressiveness.

The usefulness of realistic case examples has often been valued. Commonly, it is found that the use of realistic examples aids the learning of university students; a finding we also published in previous works [51], [52]. This also applies to the use of industry examples in an industrial setting. Our findings in this paper go even further, by stating that industry professionals benefit from the controlled application of the instructed material in a real project from their day-to-day work, which aligns with the findings by Karasneh, Jolak & Chaudron [53] who showed with a controlled experiment that the use of real examples is beneficial.

## V. Discussion

First, we will discuss the limitations and threats to validity of this experience report. Second, based on our findings and under consideration of the limitations and findings from other researchers, we will derive general guidelines.

### A. Threats to Validity

Findings result from the application of the online course within one company unit of the Schaeffler AG that was selected for the course. The company unit was selected by the company's management to provide realistic and representative insights. However, our findings might not be generalizable to other companies. The foundations for the designed teaching material have been developed in close collaboration with other industrial partners from automotive, avionic and industry automation domains. Hence, we are confident that the findings are not limited to the particular industry unit. The majority of involved industry partners must be considered large, internationally operating, Germany-based companies, which might limit transferability to small and medium-sized enterprises (SME). However, from working with some SME to develop, improve, and apply the model-based engineering methodology, we can assume that the basic approach also holds for SMEs. The approach has shown to be a good fit for, e.g., a SME in the healthcare domain as well as a SME in the robotics domain. Nevertheless, although the model-based development approach fits the needs of these companies, it must be said that typically the requirements engineering part is not as extensively elaborated in SME as it is in large companies. While a thorough requirements engineering approach is highly appreciated by these SME, we assume that for teaching requirements engineering, an approach placing more emphasis on the basic concepts of requirements engineering is needed as the personnel involved often do not have the vast amount of experience as professionals from larger companies have.

Comparison of our results with findings from other researchers shows that our results are comparable. First, this substantiates our claim that we assume our findings to be generalizable to a certain degree. Second, our findings substantiate findings from other researchers, which also have only been derived from single case study research setups. Under consideration of these circumstances, we assume that our findings are not only sufficient to highlight the problems that exist when designing online courses for industry settings but that our proposed pieces of advice for designing model-based requirements engineering courses are also useful for designing such courses. Therefore, we are confident in deriving following guidelines in the synthesis of our findings.

### B. Findings and Guidelines

As shown in Section III, our observations lead to several findings regarding the teaching of model-based requirements engineering to industry professionals. Based on these findings and under consideration of the threats to validity and the findings of other authors from the related work, we can derive some general guidelines for teaching model-based requirements engineering to industry:

*1) Teach Requirements Engineering with Models, Not Models for Requirements Engineering:* Initially companies stressed the need to limit the time for instruction. Particularly, it was stated that participants will already have taken part in other in-house education programs and can, hence, be considered proficient in UML and SysML. When starting training, it became quickly clear that participants knew the modeling languages and how to use them, e.g., to describe code on an abstract level or to sketch an architecture. However, using models to foster requirements engineering is different. The participants struggled with what to document. Particularly, the participants were eager to create detailed, very technically advanced specification models, which fail to answer why a requirement exist (i.e. do not reveal the source and the intention of the requirement but focus on technical implementation details). Another problem was, while all participants were aware what requirements are, how they are documented in the company, and how to do this (which actually resulted in requirements specifications of decent quality), the participants, usually not mainly entrusted with requirements engineering tasks, did not know why they should document requirements, why it is important to make things explicit, which have been known and were done this way for years, and what a good requirement should look like (e.g., self-contained, unambiguous, atomic). In addition, repetition of requirements engineering theory was desired to refresh participants' memory and to have all necessary information available.

Hence, good model-based requirements engineering cannot be taught without teaching requirements engineering. Consequently, a revision of the course placed more emphasis on good requirements engineering practices in general and on how to follow these practices when doing model-based requirements engineering. Using the umbrella of teaching model-based requirements engineering also had the positive side-effect that learners must not be told what they did for years was wrong, which might be offensive to some learners. Instead, under the umbrella of model-based requirements engineering, good requirements engineering practices could be taught by saying, e.g., look when we are dealing with models, it is important to note the source of a requirement, and we can do this quite easily using meta data or traceability links.

So, teaching model-based requirements engineering needs to teach good requirements engineering practices as well. However, giving participants the impression that first and foremost model-usage is taught increases the acceptance of the course by the learner, which is a great advantage.

*2) Place Emphasis on the Why and on the Benefits:* We observed that industry professionals easily grasp how things work. For instance, how a goal model can be created, how a goal model can be revised, and how a goal model can be traced to other requirements artifacts. However, we experienced that it was often unclear to participants why they should do this. For instance, in the application project certain measurements that must be adhered to are defined in the requirements. In some cases, it is a radius that must not be larger than a particular length or in other cases an angle that must be lower than a certain degree. It turned out that all these different measurements result from the same law, which limits the distance a parking car is allowed to move on its own. All angle and length values express how this particular distance has an impact at different places of the product. As it was abundantly clear to industry participants that this was the case, they did not see the need to document the original distance a parking car is allowed to move; although it is needed to document that this is required by law, to allow for easy access to the requirements specification by non-technical personnel and stakeholders. However, after discussing this issue, we became aware that this was not an attempt to protect their individual knowledge but rather that they did not know what the benefits of documenting something as obvious (to them) as this, are. Discussing these benefits, e.g., for re-use in the next project, or for change management if the law changes, participants quickly expressed that they very much liked the idea as this would make their jobs easier.

*3) Beware of Modeling Languages Already Known:* Models in requirements engineering fulfill many purposes, they support the elicitation and negotiation of requirements, and, of course, the unambiguous specification of requirements. The latter is quite intuitive for learners. Hence, models are mostly created with this purpose in mind. This often leads to very fine-grained specifications of very technical aspects. However, those specifications are commonly not preferable for involving stakeholders not familiar with technical details. This effect is heightened when it comes to modeling languages already known by the participants. Particularly, we experienced that this was the case for context models, scenario models, and system requirements models.

For instance, in context modeling, we used SysML block definition diagrams, SysML activity diagrams and SysML state machine diagrams. The participants were quite familiar with the modeling languages. This resulted in learners not paying much attention to the instruction material, as participants assumed they would already know everything needed. However, during application of the modeling techniques to the product, it turned out that the models created were not requirements models but rather models used for specifying the architecture of software. Hence, extra effort had to be expended to teach participants the open-minded attitude of requirements engineering and why it is important to not forestall design decisions during requirements engineering.

For another example, in scenario modeling, we used ITU message sequence charts for documenting the scenarios, which is beneficial as high-level message sequence charts can be used to give an additional overall structure to the single scenarios. While this worked better than the use of UML diagrams and participants got familiar with the use of scenario models in requirements engineering, this was hampered when it came to applying scenario modeling in the actual work environment. The company-wide used modeling tool did only support UML sequence diagrams, hence, participants were unable to use all concepts of ITU MSC and, again, fell back into the way they used to model diagrams with the tool.

We found the use of goal models optimal for the introduction of requirements engineering and model use in requirements engineering as goal models were completely unknown to the participants. This allowed participants to be open minded so that the use of these models for requirements engineering purposes could be taught. In contrast to the other modeling languages mentioned above, we could not identify effects of reusing knowledge in a manner that hampers requirements engineering. It might be interesting to investigate whether other non-UML modeling languages can be used similarly well in teaching mode-based requirements engineering to professionals.

*4) Pay Attention to Over and Under Specification:* Closely related to the aforementioned advice about being careful with using modeling languages already known by the learner is the need to pay attention to over and under specifications. We noticed a tendency for both while teaching model-based engineering to industry professionals. On the one hand, participants often documented the solution right away, not thinking about the problem which might not necessarily be solved this way. On the other hand, when participants were made aware of this shortcoming, their models tended to become too sketchy. Hence, we found it to be of particular importance to teach participants that we do not forestall design decisions during requirements engineering but that we place particular emphasis on defining the problem world at a detailed level.

*5) Do Not Disregard Natural Language Requirements:* Even when using model-based requirements engineering, some requirements will still have to be documented in natural language, e.g., requirements which are relevant as legally binding documents. Hence, participants need to not only learn what good practices for model-based requirements engineering are, but, furthermore, what good natural language requirements look like. We found goal modeling to be very beneficial when it comes to linking natural language requirements to model-based requirements, as goals are very close to natural language requirements. Specifically, goals can be used to close the gap between natural language requirements and model-based requirements by linking each natural language requirement to one goal. As goals are often seen as one of the most important outcomes of requirements engineering [22] and can be used for stakeholder negotiations and for verification and testing purposes in the later stages, it is of importance to specify goals of high quality. For this, knowledge about documenting natural language requirements well is beneficial.

*6) Teach Traceability:* In model-based requirements engineering multiple models are created. For instance, to investigate different alternatives, to build different views on the system or the system's context, or to separate the intentions of different stakeholders. These artifacts are all related in some way or other. Learners should, thus, not only learn why it is beneficial to create such a multitude of diagrams and models, but also how these relate to each other, how to document relationships, how to detect inconsistencies, etc. While we initially focused on teaching the individual models separately to allow for short self-contained learning blocks, we identified this approach as problematic as the relationships between the different models and the big picture gets blocked. Therefore, we consider it very helpful to teach traceability concepts right from the beginning. Teaching traceability and showing how traceability can be instantiated between different models helps not only understand traceability as a concept, but also the relation between the different models created during model-based requirements engineering. For a good start, teaching about the relation between goal and scenario models has proven beneficial. We experienced during several coaching sessions that the additional use of a use case grouping all scenarios related to a certain goal can further improve the overall comprehensibility of the relationships between goals and scenarios.

*7) Ensure Appropriate Tool Support for a Teaching Environment:* When considering tool support, it is important to distinguish between the need for tool support for the actual work environment and tool support for a teaching environment. Right away from the beginning, i.e., during the conception of the model-based requirements engineering course industry partners stressed the need for adequate tool support. This need was again expressed by the actual participants. However, we experienced that, the company desires the use of the actual work environment tool in teaching. While this is important from a management point of view, it turned out that this was not necessarily the case for the participants. It was revealed that this might not be the best option for two reasons: complexity and adherence to instructed material.

The tool used in the company can be considered quite complex, as it is the one modeling tool everyone uses for every kind of modeling task. Hence, the tool is very powerful, but also very complex. Consequently, you might end up rather teaching the tool than the method. Furthermore, the tool used in the workplace was not ready for all modeling languages needed. Moreover, the tool was unable to cope with uncertainties which are common in early requirements engineering phases. For instance, modeling a sequence diagram required specifying every detail including defining formal conditions right away. Hence, models could often not be created as would have been needed for requirements engineering purposes.

To avoid such issues, we found the use of a simplified tool very valuable for teaching model-based requirements engineering. Industry professionals were not bothered by the limitations of this tool and were aware that there is a difference to the one used in the work environment. Actually, it was appreciated that the tool was close to the instructed material and, therefore, better supported learning . However, even if a simplified tool support is used, which is tailored for teaching, at some point the need emerges to train the participants with the actual work environment tool setup to aid participants in transferring their knowledge into their day-to-day work. This means that you also might need to have answers and workarounds for the shortcomings of the actual tool. Nevertheless, we experienced that learners preferred learning the modeling languages and associated methods first and using the actual tool second. Particularly, we recognized that participants could often easily adapt the instructed material and came up with suitable workaround solutions on their own.

*8) Support Learners in Applying the Newly Learned Techniques in Real World Settings:* While it is of importance to teach good requirements engineering practices and the concepts of model-based requirements engineering as well as the different modeling languages used upfront, we found it also very helpful to accompany participants during application of the instructed material in a real-world development project. This helped us tremendously in understanding the problems of the industry participants. However, it is not a good idea to only rely on the application of the material in the company. Thorough upfront teaching is needed as otherwise industry professionals will also struggle with the application of the simple aspects of model-based requirements engineering. This way, the interaction between teacher and learner regarding the real-world application can be directed to the most difficult parts of requirements engineering education.

## VI. CONCLUSION

Model-based requirements engineering techniques have been proposed by researchers on a regular basis. Approaches place either emphasis on the use of certain models to foster, for instance, elicitation or negotiation or on conducting model-based requirements engineering in a complete model-based fashion. The latter, among others, supports the close integration with model-based systems and software engineering in general and therefore brings several benefits such as the improved traceability between requirements artifacts and design artifacts. However, to introduce model-based techniques in industry, industrial training is indispensable.

In this paper, we reported our experiences as educators and learners from designing and applying an industrial training program to train industry professionals in model-based requirements engineering. The training program includes an online course for model-based requirements engineering accompanied by webinars and workshops to allow for learner and teacher interaction.

The training program was designed in cooperation with industry partners, conducted in industry, and then the instructed material was also applied by learners to their day-to-day work as part of the training program. This allowed us to gain insights into industry needs when it comes to industrial training of model-based requirements engineering. In this paper, we reported our findings and gave advice on important aspects to consider when teaching model-based requirements engineering to industry professionals. Among others, we stressed the need to place emphasis on good requirements engineering practices as well and not only focus on model-orientation. From our experience, this can be aligned well with teaching model use in requirements engineering, as we noticed that many general requirements engineering tasks are learned better when taught alongside model-based requirements engineering than when taught in isolation.

Our findings align with phenomena commonly reported in the related work: the need for more and better industrial training (e.g., to foster technology transfer from academia to industry, to make industrial development processes fit for future) and the increasing desire for online courses (e.g., to improve and to diversify university education, to design massive open online courses, to allow for efficient industrial training reducing the amount of time needed for instructing material). This in combination with our findings allowed us to propose guidelines on teaching model-based requirements engineering to industry professionals.

## ACKNOWLEDGMENT

This research has partly been funded by the German Federal Ministry for Education and Research (BMBF) under grant no. 01IS15058C and 01IS15058G.